\documentstyle[12pt]{article}
\def\double{\baselineskip 24pt \lineskip 10pt}
\begin{document}
  \begin{titlepage}
  \begin{center}
\vspace{.2cm}
\Large
{\bf Time-Symmetrization and Isotropization of Stiff-Fluid Kantowski-Sachs
     Universes\\}
\vspace{0.5cm}
\normalsize
\large{Mariusz P. D\c{a}browski}\\
\normalsize
\vspace{1.0cm}
{\em Institute of Physics, University of Szczecin, \\ Wielkopolska 15,
70-451 Szczecin, Poland}\\
\end{center}
\vspace{.2cm}
\baselineskip=24pt
\begin{abstract}
\noindent
\double

It is shown that growing-entropy stiff-fluid Kantowski-Sachs universes become
time-symmetric (if they start with time-asymmetric phase) and isotropize.
Isotropization happens without any inflationary era during the evolution
since there is no cosmological term here. It seems that this approach is an
alternative to inflation since the universe gets bigger and bigger
approaching 'flatness'.

\end{abstract}
\vspace{1.0cm}
{\small short title: time-symmetrization and isotropization}

\end{titlepage}

\vspace{.6cm}
\section{INTRODUCTION}
\vspace{.6cm}
\double

Recently, we have considered some properties of cyclic closed universes under
the assumption that the total entropy of the universe increases from cycle to
cycle \cite{BD}. The growing entropy universes were first investigated by
Tolman \cite{TOL} and developed qualitatively by Zeldovich and Novikov
\cite{ZN}.
The result is that the entropy growth forces the universe to become bigger and
bigger in size and its total volume grows, making the universe 'flat'. It seems
that it is a kind of non-inflationary solution of the standard cosmology
problem \cite{GU}.

It has been shown that positive cosmological constant Kantowski-Sachs universes
admit the inflationary phase during which they isotropize, i.e., the shear
decays
exponentially towards zero \cite{GR,GRE}. On the other hand, even some
inhomogeneous
models homogenize during the positive cosmological-constant-driven inflation
\cite{TOM}.
So, in general, the universe can start with a very inhomogeneous and
anisotropic
phase, then goes through the inflationary era, subjecting homogenization and
isotropization. Finally, it becomes friedmannian and still it can be blown in
order to approach 'flatness' \cite{GU}.

In this paper we deal with growing entropy stiff-fluid anisotropic
Kantowski-Sachs universes. The reason is that as we have shown before \cite{BD}
the dust-filled Kantowski-Sachs universes do not isotropize, violating the
general
picture for a couple of models. We will prove that the isotropization is the
case for stiff-fluid Kantowski-Sachs models. We have not considered any exact
inhomogeneous models so far.

\vspace{.6cm}
\section{GROWING ENTROPY UNIVERSES}
\vspace{.6cm}
\double

The Kantowski-Sachs metric reads as \cite{KS,WE1,WE2}
\begin{equation}
ds^2 = dt^2 - X^2(t)dr^2 - Y^2(t) \left[ d \theta^2 + \sin^2{ \theta}d \phi^2
\right]   ,
\end{equation}
where $X$ and $Y$ are the scale factors, and the field equations for closed
models are ($p$ is the pressure and $\varrho$ is the energy density)
\begin{eqnarray}
2 \frac{\dot{X}\dot{Y}}{XY} + \frac{1 + \dot{Y}^2}{Y^2} = \varrho   ,\\
2 \frac{\ddot{Y}}{Y} + \frac{1 + \dot{Y}^2}{Y^2} = - p   ,\\
\frac{\ddot{X}}{X} + \frac{\ddot{Y}}{Y} + \frac{\dot{X}\dot{Y}}{XY} = - p   .
\end{eqnarray}

For stiff-fluid source of matter $p = \varrho$, and the second law of
thermodynamics for adiabatic expansion $d(\varrho V) + pdV = 0$ reads as
\begin{equation}
p = \varrho = \frac{M}{X^2Y^4} = \frac{M}{V^2}   ,
\end{equation}
with $M =$ const., and $V$ is the volume. One can easily show that the
constant $M$ is strictly related to the growth of entropy. From the second law
of thermodynamics we have $TdS = d(\varrho V) + pdV$. The Eq. (5) can be
rewritten to give $\varrho V = M/V$ and we may put this into the second law to
obtain $TdS = d(M/V) + pdV = dM/V$. Since both $T$ and $V$ are positive, then
for increasing values of the constant $M$ the entropy grows as well. However,
in such a case we deal with stiff-fluid particles entropy rather than with
standard dust particles or radiation entropy. Also,
in our approach there is no entropy growth within cycles. The idea is to add
some entropy at the beginning of each cycle. In fact, the results of this
approach are similar to the standard imperfect fluid approach admitting bulk
viscosity \cite{MU,HE1,HE2}, but the mathematics is not so complicated. The
parametric solution of the field equations (2)-(4) is \cite{MW}
\begin{eqnarray}
X(\eta) & = & b \left( \left| \tan{ \left( \eta + \eta_{0} \right)} \right|
 \right)^
{\pm \sqrt{1 - \frac{M}{a^2}}}   ,\\
Y(\eta) & = & \frac{a}{b} \sin{2 \left( \eta + \eta_{0} \right)} \left( \left|
\tan{ \left( \eta + \eta_{0} \right)} \right| \right)^{\pm \sqrt{1 -
\frac{M}{a^2}}}   ,\\
t(\eta) & = & 2 \int Y(\eta) d \eta   ,
\end{eqnarray}
and $a, b, \eta_{0}$ are constants. Let us define a new constant
\begin{equation}
c = \sqrt{1 - \frac{M}{a^2}}   ,
\end{equation}
and
\begin{eqnarray}
0 \leq c < 1  \nonumber
\end{eqnarray}
necessarily, which means that the constants $M$ and $a$ are related by the
condition (i.e., the growth of entropy is limited)
\begin{eqnarray}
a^2 \geq M   \nonumber   .
\end{eqnarray}
If $c = 0$ ($M = a^2$) the models are time-symmetric and both the initial and
final singularities are the so-called barrel singularities \cite{COL} (i.e. $X
= b =$
 const. and $Y = 0$ at them). If $c \neq 0$ the models do not possess
time-symmetry and there is one point singularity $X = Y = 0$ and one cigar
singularity $Y = 0, X = \infty$.

It seems that due to an absolute value of $\tan{(\eta + \eta_{0})}$ which
appears in (6)-(7), the models cannot be considered as passing through
one to another cycle. It is because we cannot connect suitable values of
$\sin{(\ldots)}$ and $\tan{(\ldots)}$ in different consecutive ranges of
$\eta$ , unless we take negative values of the constant $a$ in negative
$Y(\eta)$ cycles.
%(Figs. 1a-b)
 The other possibility to connect different
cycles is to take only the time-symmetric case $c = 0$ into account, but
as we shall see it is not as interesting as the time-asymmetric case.
If $c = 0$, then from (6)-(8) we have
\begin{eqnarray}
X(\eta) & = & b   ,\\
Y(\eta) & = & \pm \frac{M^{\frac{1}{2}}}{b} \sin{2 \left( \eta + \eta_{0}
\right)}   ,\\
t(\eta) & = & t_{0} \pm \frac{M^{\frac{1}{2}}}{b} \cos{2 \left( \eta + \eta_{0}
\right)}   ,
\end{eqnarray}
where the plus sign refers to odd cycles and the minus sign refers to even
cycles. It is easy to notice that Eqs. (10)-(12) can be deparametrized to give
\begin{eqnarray}
X_{n}(t) & = & b     ,\\
Y_{n}(t) & = & \frac{M^{\frac{1}{2}}}{b} \sqrt{ 1 - \frac{b^2}{M}
(t - t_{0n})^2}      ,
\end{eqnarray}
where the index $n$ refers to the n-th cycle with a suitable constant
$t_{0n}$.
%(Fig.2)
Assuming that the entropy grows in each cycle of a factor $\omega$, i.e.,
\begin{equation}
M_{n+1} = \omega M_{n}      \left( \omega > 1 \right)     ,
\end{equation}
one can calculate that
\begin{equation}
t_{n} = 2 \frac{M^{\frac{1}{2}}}{b} \sum_{i=1}^{n} \omega^{\frac{i-1}{2}}
\end{equation}
is the cosmic time at the end of the n-th cycle,
\begin{equation}
Y_{n,max} = \omega^{\frac{n-1}{2}} \frac{M_{1}^{\frac{1}{2}}}{b}
\end{equation}
is the maximum value of the scale factor Y,
\begin{equation}
t_{n,max} = 2 \frac{M^{\frac{1}{2}}}{b} \left[ \left( \sum_{i=1}^{n-1}
\omega^{\frac{i-1}{2}}\right) + \frac{\omega^{\frac{n-1}{2}}}{2} \right]
\end{equation}
is the moment of maximum expansion, while
\begin{equation}
t_{0n} = 2 \frac{M^{\frac{1}{2}}}{b} \left[ \left( \sum_{i=1}^{n-1}
\omega^{\frac{i-1}{2}}\right) + \frac{\omega^{\frac{n-1}{2}}}{2} \right]
\end{equation}
fixes the constant within the n-th cycle.

The volume in the $(n+1)$-th cycle is
\begin{equation}
V_{n+1}(t) = X_{n+1}Y_{n+1}^2 = \frac{M_{n+1}}{b} \left[ 1 -
\frac{b^2}{M_{n+1}}
(t - t_{0n+1})^2 \right]    ,
\end{equation}
so it grows in subsequent cycles and the energy density, according to (5),
\begin{equation}
\varrho_{n+1}(t) = \frac{M_{n+1}}{V_{n+1}^2} = \frac{b^2}{M_{n+1}}
\left[ 1 - \frac{b^2}{M_{n+1}} (t - t_{0n+1})^2 \right]^{-2}
\end{equation}
decreases, similarly as in the Friedmann universes. The anisotropy for the
time-symmetric case $(c = 0)$ is
\begin{equation}
\sigma_{n+1}(t) = \mp \frac{1}{\sqrt{3}} \frac{\dot{Y}_{n+1}}{Y_{n+1}} =
\mp \frac{b^2}{\sqrt{3}M_{n+1}} \frac{t - t_{0n+1}}{ 1 - \frac{b^2}{M_{n+1}}
(t - t_{0n+1})^2}    ,
\end{equation}
which in the limit $M_{n+1} \rightarrow \infty$ gives
\begin{equation}
\sigma_{n+1}(t) \approx \mp \frac{b^2}{\sqrt{3}M_{n+1}} \left( t - t_{0n+1}
\right)   .
\end{equation}
This means that the anisotropy decreases in subsequent cycles and the same
does the expansion
\begin{equation}
\Theta_{n+1}(t) = \pm 2 \frac{\dot{Y}_{n+1}}{Y_{n+1}} \approx
\frac{b^2}{\sqrt{3}M_{n+1}} \left( t - t_{0n+1} \right)     .
\end{equation}
{}From (24)-(25) it follows that the shear-to-expansion rate is constant
\begin{equation}
\frac{\sigma}{\Theta} = \frac{1}{2\sqrt{3}}   .
\end{equation}
In the case $c \neq 0$, we can think about matching different cycles with no
regard to the fact that every second cycle will not be similar and that the
following
cycles will be the mirror images of the previous ones. However, the basic
feature
seems to be consistent. Whenever the final singularity is a cigar, the
following
initial singularity is  a cigar as well. Whenever the final singularity is
a point, the following initial singularity is a point either.

First, let us take the upper signs in powers of (6)-(8), and choose the
constant
a such that $a > 0$ for odd cycles and $a < 0$ for even cycles. Then in the
first
cycle we have
\begin{eqnarray}
X_{1}(\eta) & = & b \left( \left| \tan{ \left( \eta + \eta_{0} \right)}
\right| \right)^{c_{1}}   ,\\
Y_{1}(\eta) & = & \frac{a}{b} \sin{2 \left( \eta + \eta_{0} \right)} \left(
 \left|
\tan{ \left( \eta + \eta_{0} \right)} \right| \right)^{- c_{1}}   ,\\
t_{1}(\eta) & = & 2 \int Y_{1}(\eta) d \eta   ,
\end{eqnarray}
with
\begin{equation}
c_{1} = \sqrt{1 - \frac{M_{1}}{a^2}} > 0   ,
\end{equation}
so, in the second cycle
\begin{eqnarray}
X_{2}(\eta) & = & b \left( \left| \tan{ \left( \eta + \eta_{0} \right)}
\right| \right)^{c_{2}}   ,\\
Y_{2}(\eta) & = & \frac{a}{b} \sin{2 \left( \eta + \eta_{0} \right)} \left(
 \left|
\tan{ \left( \eta + \eta_{0} \right)} \right| \right)^{- c_{2}}   ,\\
t_{2}(\eta) & = & 2 \int Y_{1}(\eta) d \eta   ,
\end{eqnarray}
and
\begin{equation}
c_{2} = \sqrt{1 - \frac{M_{2}}{a^2}} > 0   ,
\end{equation}
where accorrding to (9)
\begin{eqnarray}
c_{2} < c_{1}   \nonumber   ,
\end{eqnarray}
which means that values of $X(\eta)$ increase together with $M$ in each
subsequent cycle in the interval $0 \leq X < b$ and decrease in the interval
$b < X < \infty$. In the limit of large $M \rightarrow a^2$ the values of $X$
becomes constant and the model reaches the time-symmetric case $c = 0$.
%(Fig.3)
If $c < 0$ the situation
changes in the way that the first cycle begins with a cigar singularity
$X = \infty, Y = 0$ instead of a point singularity $X = Y = 0$, i.e., the whole
picture translates of $\frac{\pi}{2}$ but the physical properties remain the
same.
On the other hand, if the constant $M$ grows, $Y(\eta)$
increases in each subsequent cycle in the ranges where its values are smaller
(i.e., where $0 < Y < \frac{a}{b}$), and decreases in the ranges where it has
maxima
and its values are larger (i.e., where $Y > \frac{a}{b}$), giving finally
time-symmetric picture with $c = 0$.
% (Fig.4)

 From the above we can conclude a very important remark that the growth of
entropy forces the universe to become time-symmetric, even if it was not like
that at the beginning.

The volume in the $(n+1)$-th cycle is ($\eta_{0} = 0$, $c_{n+1}$ is given by
(9) with $M_{n+1}$)
\begin{equation}
V_{n+1}(\eta) = X_{n+1}Y_{n+1}^2 = \frac{a^2}{b} \sin^2{2\eta}
\left( \left| \tan{\eta}\right| \right)^{- c_{n+1}}   \nonumber   ,
\end{equation}
and it behaves very similar to the behavior of $Y(\eta)$.
%(i.e. decreases wherever it possesses a maximum and grows, if it does not)
%%until
%it reaches the value given by (18) for $c = 0$ and it starts growing with the
%entropy growth (time-symmetric case).
The anisotropy
\begin{equation}
\sigma_{n+1}(\eta) = \pm \frac{1}{\sqrt{3}} \frac{1}{2Y} \left(
\frac{X_{,\eta}}{X} - \frac{Y_{,\eta}}{Y} \right) = \pm \frac{b}{a\sqrt{3}}
\frac{\left|\tan{\eta}\right|^{c_{n+1}}}
{\sin{2\eta}} \left[ \frac{c_{n+1}}{\cos^2{\eta} \left|\tan{\eta}\right|} -
\cot{2 \eta} \right]   \nonumber
\end{equation}
decreases consecutively because of the decrease of $c$, until it reaches $c =
0$
, and still decrease in the time-symmetric case (22). Finally, the
expansion
\begin{equation}
\Theta_{n+1}(\eta) = \pm \frac{1}{2Y} \left(
\frac{X_{,\eta}}{X} - \frac{Y_{,\eta}}{Y} \right) =
\pm 2 \frac{b}{a} \left| \tan{\eta} \right|^{c_{n+1}} \frac{\cos
{2\eta}}{\sin^2{2\eta}}   \nonumber
\end{equation}
always decreases.

Of course the discussion of the time-asymmetric case was carried out in
conformal time $\eta$ and the
investigation of the problem in cosmic time would require the exact
integration of (8). However, from the behaviour of the time-symmetric case
(13)-(14) it seems that the general picture will not be changed.

\vspace{.6cm}
\section{CONCLUSIONS}
\vspace{.6cm}

We have shown that, unlike the dust-filled models \cite{BD}, the stiff-fluid
Kantowski-Sachs cyclic universes isotropize under the assumption that the
entropy grows from cycle to cycle. Also, we have proved that time-asymmetry of
these models gradually vanishes and the models finally become time-symmetric.

\begin{center}
{\bf Acknowledgments}
\end{center}
\vspace{.6cm}
The author wishes to thank John Barrow for helpful discussions. A part of this
work was performed during the author's stay at the Astronomy Centre, University
of Sussex, UK.

\pagebreak

%\pagebreak
%\begin{center}
%{\bf Figure Captions}
%\end{center}

%Fig.1.a,b. The behaviour of the scale factors $X$ and $Y$ (6)-(7) in a
%stiff-fluid Kantowski-Sachs model without the entropy growth. The values of
%the parameter $\eta$ are given in radians and the constant $c$ is taken
%to be positive.

%Fig.2. The behaviour of the scale factors $X$ and $Y$ (13)-(14) in a
%time-symmetric stiff-fluid Kantowski-Sachs model with growing entropy.
%The values of the parameter $\eta$ are given in radians ($c$ positive).

%Fig.3. The behavior of the scale factor $X$ in a ti\-me-a\-sym\-met\-ric
%stiff-fluid Kan\-to\-wski-Sachs model with growing entropy.

%Fig.4. The behavior of the scale factor $Y$ in a ti\-me-a\-sym\-met\-ric
%stiff-fluid Kan\-to\-wski-Sachs model with growing entropy.


\frenchspacing
\begin{thebibliography}{99}
\bibitem{BD} Barrow J D and D\c{a}browski M P 1995 {\it Mon. Not. R. Astr.
      Soc.} {\bf 275} 850 (1995)
\bibitem{TOL} Tolman R C 1934 {\it Relativity, Thermodynamics and Cosmology}
      (Oxford: Clarendon Press)
\bibitem{ZN} Zeldovich Ya B and Novikov I 1983 {\it The Structure and Evolution
of the
      Universe} (Chicago: The University of Chicago Press)
\bibitem{GU} Guth A 1981 {\it Phys. Rev.} D {\bf 23} 347
\bibitem{GR} Gr{\o}n \O 1986 {\it Journ. Math. Phys.} {\bf 27} 1490
\bibitem{GRE} Gr{\o}n \O and Ericsen E 1987 {\it Phys. Lett.} {\bf A121} 217
\bibitem{TOM} Tomita K 1994 {\it Phys. Rev.} {\bf D48} 5634
\bibitem{KS} Kantowski R and Sachs R K 1966 {\it Journ. Math. Phys.} {\bf 7}
443
\bibitem{WE1} Weber E 1984 {\it Journ. Math. Phys.} {\bf 25} 3279
\bibitem{WE2} Weber E 1985 {\it Journ. Math. Phys.} {\bf 26} 1308
\bibitem{MU} Murphy G L 1973 {\it Phys. Rev.} D {\bf 8} 4231
\bibitem{HE1} Heller M, Klimek Z and Suszycki L 1973 {\it Astroph. Space Sci.}
      {\bf 20} 205
\bibitem{HE2} Heller M and Szyd{\l }owski M 1983 {\it Astroph. Space Sci.}
      {\bf 90} 327
\bibitem{MW} Mim\'{o}so J and Wands D G 1995 {\it Preprint} gr-qc/9501039
\bibitem{COL} Collins C B 1977 {\it Journ. Math. Phys.} {\bf 18} 2116

\end{thebibliography}
\end{document}